\documentstyle[12pt]{article}

\font\sqi=cmssq8
\def\DR{\rm I\kern-1.45pt\rm R}
\def\DC{\kern2pt {\hbox{\sqi I}}\kern-4.2pt\rm C}

\begin{document}
\thispagestyle{empty}
\begin{flushright}
Preprint JINR E2-95-383\\
hep-th/9508137\\
\end{flushright}
\vspace{2cm}
\begin{center}
{\large\bf Quantum oscillator and a bound system of two dyons}\\
\vspace{2cm}
 {\large V. M. Ter--Antonyan
\footnote{e-mail:terant@thsun1.jinr.dubna.su},  A. Nersessian
\footnote{e-mail:nerses@thsun1.jinr.dubna.su}}

\vspace{1cm}

{\it Bogoliubov Laboratory of Theoretical Physics,\\
Joint Institute for Nuclear Research,\\
Dubna, Moscow Region, 141980, Russia}

\end{center}

\begin{abstract}

It is shown that $U(1)$--Hamiltonian reduction of a
four--dimensional isotropic quantum
oscillator results in a bound system of two spinless Schwinger's dyons.
Its wavefunctions and spectrum are constructed.
\end{abstract}
\newpage
 {\bf Introduction  }\\

In our  paper \cite{mpl}  a classical integrable
bound  system describing the non-relativistic interaction of a
spinless particle with the $U(1)$-dyon in the
centre-of-mass system (``charge-dyon" system)  was constructed by the
$U(1)$ Hamiltonian reduction of an oscillator on the twistor space.
 This system is specified by hidden symmetry caused by
 the Runge--Lenz vector-- like constants of motion.
It describes not only the ``charge-dyon" interaction,
 but also the ``dyon-dyon" one.

In the present paper,  by an analogous  reduction
at the {\it quantum} level, we construct the Schr\"odinger
equation, constants of motion and wave functions and spectrum
for a non-relativistic bound system  describing the interaction
of two Schwinger dyons in the center--of--mass system.
Despite simplicity, this system inherits such
remarkable properties of  monopole systems
(see, e.g. \cite{J} and refs therein), as a degenerate ground state
 and ``spin transmutation".
A quantum system, unlike from a classical one,
explicitly depends from the vector potential of a monopole field.
It was shown in the Zwanziger paper \cite{Z} that only the system,
 specified by the  vector potential of the Schwinger monopole,
 can be interpreted as the ``dyon-dyon" one, while similar system
with the vector potential of the Dirac  monopole characterizes only
the ``charge-dyon" interaction.
Notice that such a quantum system , depending
on the Dirac monopole vector potential has been constructed by a
similar reduction in   ref. \cite{qMIC}. However, the authors did not know
the physical meaning of the system they constructed,
therefore, its important properties were unnoticed \cite{pc}.
In our consideration, we will omit the details of calculation coincident
with those of classical reduction \cite{mpl}.

We will use the following notation: $\mu$ and $\omega$ are oscillator
parameters; $z^\alpha=u^\alpha+iu^{\alpha+2}$ and
$\vec r =(x^1, x^2, x^3)$ are Cartesian coordinates of spaces
${\DC}^2=\DR^4$ and
$\DR^3$ respectively, $u=|z|, r=|{\vec r}|$;
${\vec\sigma}$ are the Pauli matrices in a standard representation;
and $a=(\frac{\mu\omega}{\hbar})^{\frac 12}$
is a parameter with the dimension of inverse length. \\

 {\bf The Schr\"rodinger equation and constants of motion }\\

An isotropic oscillator on the space ${\DC}^2$
is described by the Schr\"rodinger's
equation
\begin{equation}
\frac{\partial^2\Psi}
{{\partial z^\alpha}{\partial {\bar z}^\alpha}}+\frac{2\mu}{\hbar^2}
\left[\frac{E}{4}-\frac{\mu\omega^2 z{\bar z}}{8} \right]\Psi=0
\quad\Leftrightarrow\quad {\cal H}_{osc}\Psi=E\Psi.
\label{so}\end{equation}
Its constants of motion
\begin{equation}
{\cal J}_i =\frac{1}{2} (
 z^\alpha\sigma_{\beta\alpha}^i\frac{\partial}{\partial z^\beta}
-
{\bar z}^\alpha\sigma_{\alpha\beta}^i\frac{\partial}
{\partial{\bar z}^\beta})
, \;\;
{\cal I}_{i}=-4\frac{\hbar^2}{2\mu}\sigma^{i}_{\alpha\beta}
\frac{\partial}{\partial z^\beta}\frac{\partial}{\partial {\bar z}^\alpha}
 + \frac{\mu\omega^2}{2}\sigma^{i}_{\alpha\beta}
z^\alpha {\bar z}^\beta ,\label{Ji}
\end{equation}
form the algebra:
\begin{equation}
\{{\cal J}_k, {\cal J}_l\}=i\varepsilon_{klm}{\cal J}_m ,
\quad \{{\cal J}_k, {\cal I}_l\}=i\varepsilon_{klm}{\cal I}_m,
\quad \{{\cal I}_k ,
{\cal I}_l\}=i(2\hbar\omega)^2\varepsilon^{klm}{\cal J}_m.\label{8a}
\end{equation}
Let us now perform the quantum reduction of this system with respect to
the action of group $U(1)$ given by the operator
 \begin{equation}
{\cal J}_0 =\frac{1}{2}(
 z^\alpha\frac{\partial}{\partial z^\alpha}
-
{\bar z}^\alpha\frac{\partial}{\partial{\bar z}^\alpha}) ,\label{j0}
\end{equation}
commuting with the constants of motion (\ref{Ji})
 and with the oscillator's Hamiltonian
\begin{equation}
   [{\cal J}_0, H_{osc}]= 0,\quad
[{\cal J}_0, {\cal J}_i]=0 \quad
[{\cal J}_0, {\cal I}_i ]=0 .
\label{Jhi}\end{equation}
To this end, we introduce the operators
\begin{equation}
{\vec r} ={\bar z}^{\alpha} {\vec\sigma}_{\alpha \beta}z^{\beta},
\quad    {\hat{\vec p}}=\frac{i\hbar}{2(z\bar z)}(
 z^\alpha{\hat\sigma}_{\beta\alpha}\frac{\partial}{\partial z^\beta}
+
{\bar z}^\alpha{\hat\sigma}_{\alpha\beta}\frac{\partial}
{\partial{\bar z}^\beta})                ,
\label{xp}\end{equation}
obeying the relations
 \begin{eqnarray}
& [{\cal J}_0,{\vec r} ]=0, \quad  [{\cal J}_0, {\hat{\vec p}}]=0 &
\label{xc}\\
&  [x^k, x^l]=0,\quad  [x^{k}, {\hat p}^l ]=i\hbar\delta^{kl},\quad [{\hat
p}^k,{\hat p}^l] =
\hbar^2\varepsilon^{klm}\frac{x^m}{r^3}{\cal J}_0.
&\end{eqnarray}
To a fixed eigenvalue of the operator ${\cal J}_0  $
\begin{equation}
{\cal J}_0 \Psi= s\Psi \label{j0s}
\end{equation}
 there corresponds the wavefunction
\begin{equation}
\Psi (z,{\bar z})=\psi (\vec r) e^{is\gamma},
\label{wfs}\end{equation}
where $\gamma$ is a function conjugate to the operator  ${\cal J}_0$:
\begin{equation}
\gamma=\frac{i}{2}(\log{ z^1}/{\bar z}^1 +\log{z^2}/{\bar z}^2)
,\quad\gamma \in [0, 4\pi ):\quad  [{\cal J}_0,\gamma]=i.
\label{x}\end{equation}
Here also we have
\begin{equation}
 {\hat{\vec p}}( \psi e^{is\gamma}) =
( {\hat\pi}\psi ) e^{is\gamma} ,
\end{equation}
where
\begin{equation}
{\hat\pi}= -i\hbar\frac{\partial}{\partial{\vec r}}- {\hbar s}\vec A,
\quad
\vec A=\frac{{\vec n}_3{\vec r}}{r}\frac{\vec n_3\times\vec r}
{r^2-({\vec n_3}{\vec r})^2} .
 \label{AS}\end{equation}
and $\vec A$ is a vector potential of the Schwinger's monopole
with a unit magnetic
charge and a singular line along the axis $x^3$; $\vec n_3= (0, 0, 1)$.

Owing to the relations (\ref{Jhi}) and (\ref{xc}),
the oscillator's Hamiltonian and constants
of motion (\ref{Ji}) are expressed through
${\vec r},{\hat{\vec p}},{\cal J}_0$.
 As a result, the substitution (\ref{wfs})
reduces equation (\ref{so}) to the form
\begin{equation}
 {\hat H}\psi=-\frac{\mu\omega^2}{8}\psi,\quad
{\hat H}=\frac{\hbar^2}{2\mu} {\hat\pi}^2 -
\frac{E}{4r}+\frac{\hbar^2{s}^2}{2\mu r^2}\quad,
\label{redH}\end{equation}
whereas the constants of motion of the oscillator (2) are reduced to the
operators
  \begin{equation}
{\vec J}=-{\hat\pi}\times\vec r
 +\frac{\hbar{s}{\vec r}}{r},\quad
{\vec I}=\frac{\hbar^2}{2\mu}{\hat\pi}\times{\vec J} +
\frac{{\vec r}Е}{2r},
\label{moment}\end{equation}
i.e. to the total angular momentum of the system and to an analog of the
Runge--Lenz vector.

{}From the requirement for the wavefunction (\ref{wfs})
being single--valued we
derive immediately
\begin{equation}
s= 0,\pm 1/2, \pm 1, \ldots.
\label{dq}\end{equation}

{\it The obtained system describes nonrelativistic
interaction of two spinless dyons
with electric amd magnetic charges}  $(e_1, g_1)$ {\it and}
$(e_2, g_2)$ {\it and energy} ${\cal E}$ {\it if we put }

\begin{equation}
\frac{e_1g_2-e_2g_1}{\hbar c} =s ,\quad e_1e_2+g_1g_2=\frac{E}{4},   \quad
{\cal E} =-\frac{\mu\omega^2}{8}.
\label{parameter}\end{equation}
{\it The parameter} $\mu$ {\it represents the reduced mass;
the description holds in the centre-of-mass system } \cite{Z}.
{\it The first of formulae}
 (\ref{parameter}) {\it together with  }(\ref{dq})
  {\it acquires the meaning of
the Dirac's condition of charge quantization.}\\

{\it Remark.}
 The vector--potential shape in (\ref{AS}) depends on the choice of the
coordinate $\gamma$ conjugate to the operator ${\cal J}_0$.
For instance, the vector
potential of the Dirac's monopole
is described by the following choice of the coordinate $\gamma$
 \cite{qMIC} :
\begin{equation}
\gamma=i\log{ z^1}/{\bar z}^1 ,\gamma \in [0, 4\pi )\Rightarrow
\vec A=\frac{1}{r}\frac{\vec n_3\times\vec r}{r-({\vec n_3}{\vec r})}.
\label{AD}\end{equation}
This system can be interpreted only as the ``charge--Dirac's dyon"
 system \cite{mpl,Z}.\\

{\bf Wavefunctions and spectrum }\\

The system of equations
\begin{eqnarray}
&{\cal H}_{osc}\Psi=E\Psi&,\quad{\cal J}_i{\cal J}_i\Psi= j(j+1)\Psi,
\nonumber\\
&{\cal J}_0 \Psi= s\Psi &,\quad{\cal J}_3 \Psi= m\Psi,
 \label{j3}\end{eqnarray}
is separated in the coordinates
$u\in [0,\infty),\; \beta \in [0,\pi],
\;\alpha \in [0,2\pi), \;\gamma \in [0,4\pi)$:
\begin{equation}
z_1=u\, \cos\frac{\beta}{2}\, e^{-i\frac{\alpha+\gamma}{2}}\quad
z_2=u\, \sin\frac{\beta}{2}\, e^{ i\frac{\alpha-\gamma}{2}}.
\label{e}\end{equation}
As a result, the solution to the system (\ref{j3}) is of the form
\begin{equation}
\Psi_{Ejms}=R_{Ej}(u)D^j_{ms}(\alpha,\beta,\gamma).
\label{wfo}\end{equation}
where $D^j_{ms}(\alpha,\beta,\gamma)$
 is the Wigner  function
\begin{equation}
   D^j_{ms}(\alpha,\beta,\gamma)=e^{im\alpha}d^j_{ms}(\beta)e^{is\gamma}
,\label{d}\end{equation}
and the radial function $R_{Ej}$ obeys the equation
\begin{equation}
\frac{d^2 R_{Ej}}{d\rho^2} +\frac{3}{\rho}\frac{d R_{Ej}}{d\rho}-
[\frac{4j(j+1)}{\rho^2}+
\rho^2-\lambda ]R_{Ej}=0,
\label{rad}\end{equation}
with $\rho=(au)^2$, $\lambda =\frac{2\mu E}{\hbar^2 a^2}$, \\
 The substitution
$R_{Ej}=\rho^j{\rm e}^{-\rho /2}W(\rho)$
 reduces eq. (\ref{rad})  to the equation for the
confluent hypergeometric function
$$ \rho W''+(2j+1-\rho)W' +(\frac{\lambda}{4}-j-1)W=0.$$
The solution regular at the point $\rho =0$ is given by
$$W(\rho)=const F(j+1-\frac{\lambda}{4}, 2j+1, \rho).$$
As a result,
\begin{equation}
R_{nj}(\rho)={\rm  const}\; \rho^j {\rm e}^{-\rho/2} F(-n+1, 2j+1, \rho),
\label{wfor}\end{equation}
where $j+1-\lambda/4 =-n+1$.
Expressions  (\ref{wfo}), (\ref{d}) and (\ref{wfor})
determine the oscillator basis.

{}From the requirement $R_{Ej}(\infty)=0$
and uniqueness of the function (\ref{d})it follows, that:
 $n=1,2,3,\ldots;\;\; m,s=-j,-j+1,\ldots,j-1,j ;\;\; 2j= 0, 1,\ldots .$
Then, upon introducing the principal quantum number $N=2n+2j-2$
, we obtain the
following relations for the oscillator spectrum:
\begin{eqnarray}
&&E=\hbar\omega (N+2), \quad N=0,1,2,\ldots ;\label{eo} \\
&& 2j= 0, 1,\ldots N;\\
&& m, s=-j,-j+1,\ldots,j-1,j \;\;\;.
\end{eqnarray}
At fixed $j$  the $(2j+1)^2$ states  corresponds to the level
 $E_N$. Since $j= N/2,\;N/2-1,\ldots $, the
degree  of degeneracy of the $N$th level is equals to
$$g_N=\frac 16 (N+1)(N+2)(N+3).$$
Now we can construct the wavefunctions and spectrum of the reduced system.\\
The coordinates of space $\DC^2$
transform into the spherical coordinates of
space $\DR^3$ :$(r=u^2 ,\;\theta= \beta ,\;\phi=\alpha)$.

Comparison of (\ref{wfs}) with (\ref{wfo})  gives the
following wavefunction   of the reduced system
\begin{equation}
\psi_{njm}(\vec r ; {s}) =
{\rm const}\; R_{nj}(ar) d^j_{ms}(\theta)e^{im\phi}\label{wfd}.
\end{equation}
and expressions (\ref{parameter}), (\ref{eo}) result
in the energy spectrum for the system
\begin{equation}
 {\cal E}^s_k =
-\frac{\mu (e_1 e_2+ g_1g_2)^2}{2\hbar^2(k+|{s}|)^2},\quad  k= 1, 2 ,
\ldots\quad .
\label{senergy} \end{equation}
For fixed ${\cal E}^s_k$
$$ j= |{s}|, |{s}| +1,\ldots, k +|{s}| -1 ;\quad m =-j, -j+1,\ldots, j-1, j .
$$
Therefore, the energy levels (\ref{senergy})
are degenerated with multiplicity $g_k^s=k(k+2|s|).$

Thus, having reduced a 4--dimensional quantum oscillator, we have constructed
the Schr\"rodinger's equation for a bound system of two Schwinger's dyons, its
constants of motion, wavefunctions and the spectrum.

We stress that the quantum numbers $j, m$ characterize the total angular
momentum (spin) and its projection onto the axis $x_3$. Therefore, integer
and half--integer values of $s$ represent, respectively,  integer and half--
integer values of the system's spin .
At $s = 0$ the system becomes hydrogen--like.

Under the identical transformation
 $\phi\to\phi+2\pi$, the wavefunction of the reduced
system acquires the phase $2\pi m$:
 it is single--valued at integer $s$ and changes
in sign at half--integer $s$.

The wavefunction of the ground state ($k=1, j= |{s}|)$ of the system
 is of the form
$$
\psi_{1,m}(\vec r ;{s})={\rm const}\; r^{|{s}|} e^{-r/(|{s}|+1)}
 (\sin\theta)^{|{s}|}
(\tan\frac{\theta}{2})^{\mp m}e^{im\phi}.
$$
It is seen that the ground state is degenerated (with respect to the quantum
number $m$) and is not spherically symmetric: the system has a nonzero dipole
moment.

Note is to be made that when $m\neq\pm |{s}|$,
we have $|\psi(\theta = 0)|^2=|\psi(\theta = \pi)|^2 =0$,
which means that the system is flattened to the plane $x^3=0$
and the charge cannot be on the singular line.
This property holds valid for excited states as well.

 At $m=|{s}|$ we have $|\psi(\theta = 0)|^2\neq0, |\psi(\theta = \pi)|^2 =0$,
which implies that the charge cannot be on the lower semiaxis $x^3$.
At $m=-|{s}|$ the charge cannot be on the upper semiaxis $x^3$.\\

{\bf Acknowledgments.}The authors are thankful to L.Mardoyan for
useful discussions
and to V.I. Ogievetsky, A.I. Pashnev, A.N. Sissakian and Y.Uwano
 for interest in the work. The work  of A.N. has been made possible by a
fellowship
of the Grant No. M21300 from International
Science Foundation and INTAS Grant No. 93-2492 and have been carry out
within  the research program of the International Center of
Fundamental Physics in  Moscow.

\end{document}